\documentclass[10pt, sigconf]{acmart}

\newif\ifsubmission
\submissionfalse

\usepackage[english]{babel}
\usepackage{blindtext}
\usepackage{multirow}
\usepackage{xcolor}
\usepackage{soul}
\usepackage{array}
\usepackage{tagging}
\usepackage{graphicx}
\usepackage{caption}
\usepackage{cleveref}
\usepackage{subcaption}
\usepackage{adjustbox}
\usepackage[most]{tcolorbox}
\usepackage{float}
\usepackage[inline]{enumitem}
\usepackage{booktabs} 
\usepackage{xcolor} 
\usepackage{todonotes}
\usepackage{listings} 
\usepackage{tabularx}

\ifsubmission
\newcommand{\mcnote}[1]{}
\newcommand{\acnote}[1]{}
\newcommand{\vanote}[1]{}
\newcommand{\zqnote}[1]{}
\newcommand{\mbnote}[1]{}
\else
\newcommand{\mcnote}[1]{\todo[color=purple!40,inline]{MC: #1}}
\newcommand{\acnote}[1]{\todo[color=blue!40,inline]{AC: #1}}
\newcommand{\vanote}[1]{\todo[color=olive!40,inline]{VA: #1}}
\newcommand{\zqnote}[1]{\todo[color=red!40,inline]{ZQ: #1}}
\newcommand{\mbnote}[1]{\todo[color=green!40,inline]{MB: #1}}
\fi

\newcommand{\smartparagraph}[1]{\noindent{\bf #1}.\ }

\newcommand{\sysname}{\textsc{DMAS-Forge}\xspace}

\newcommand{\codefrag}[1]{{\ttfamily\small #1}}

\lstdefinelanguage{Go}{
  morekeywords={break,case,chan,const,continue,default,defer,else,fallthrough,
    for,func,go,goto,if,import,interface,map,package,range,return,select,
    struct,switch,type,var,nil,true,false,iota,def},
  sensitive=true,
  morecomment=[l]{//},
  morecomment=[s]{/*}{*/},
  morestring=[b]",
}

\renewcommand\footnotetextcopyrightpermission[1]{} 

\acmConference[]{1st Workshop on Systems for Agentic AI (SAA'25)}{October 13, 2025}{Seoul, Republic of Korea}

\begin{document}


\title[]{DMAS-Forge: A Framework for Transparent Deployment of AI Applications as Distributed Systems}

\author{Alessandro Cornacchia}
\affiliation{%
  \institution{KAUST}
  \country{}
}

\author{Vaastav Anand}
\affiliation{%
  \institution{MPI-SWS}
    \country{}
}

\author{Muhammad Bilal}
\affiliation{%
  \institution{KAUST}
  \country{}
}

\author{Zafar Qazi}
\affiliation{%
  \institution{LUMS \& KAUST}
  \country{}
}

\author{Marco Canini}
\affiliation{%
  \institution{KAUST}
  \country{}
}

\renewcommand{\shortauthors}{}

\crefname{figure}{Fig.}{Figs.}
\crefname{tabular}{Tab.}{Tabs.}
\crefname{table}{Tab.}{Tabs.}
\crefname{section}{\S}{\S}

\begin{abstract}
Agentic AI applications increasingly rely on multiple agents with distinct roles, specialized tools, and access to memory layers to solve complex tasks---closely resembling service-oriented architectures.
Yet, in the rapid evolving landscape of programming frameworks and new protocols, deploying and testing AI agents as distributed systems remains a daunting and labor-intensive task.
We present DMAS-Forge, a framework designed to close this gap.
DMAS-Forge decouples application logic from specific deployment choices, and aims at transparently generating the necessary glue code and configurations to spawn distributed multi-agent applications across diverse deployment scenarios with minimal manual effort. 
We present our vision, design principles, and a prototype of DMAS-Forge. Finally, we discuss the opportunities and future work for our approach.
\end{abstract}

\maketitle

\section{Agentic AI}\label{sec:agent}
Agentic AI represents the next stage in the evolution of intelligent systems. Agentic AI augments a traditional AI model by incorporating advanced capabilities such as planning, reasoning, contextual memory, and tool use.
AI agents can dynamically direct their own tool usage and follow the set of steps towards achieving a goal~\cite{yao2023react,langgraph-workflows}.
These features enable agents to work autonomously with minimal human intervention. Therefore, AI agents are designed to operate in dynamic environments where adaptability and strategic decision-making are essential. 

\smartparagraph{Multi-agent systems (MAS)} MAS extend this paradigm by enabling multiple AI agents to collaborate in pursuit of shared objectives. Each agent within such a system possesses a degree of autonomy, specialized skills, and a localized view of the broader environment. Through coordination, communication, and task-sharing, MAS can address problems that are too large, complex, or interdependent for a single agent to manage effectively.
Attempting to assign a highly complex task to a single agent often leads to challenges: instructions may become overly complicated, the likelihood of errors increases, validation becomes more difficult, and the agent may require excessive access to tools and permissions. By distributing responsibilities across multiple agents, MAS reduces these risks. Each agent can be restricted to a well-defined scope of action, equipped with the tools necessary for its role, and powered by the most appropriate AI models for its specialized function. Furthermore, the use of distinct memory systems across agents enhances adaptability, as agents can draw on task-specific knowledge while contributing to a collective goal.
Therefore, while an individual AI agent can perform a wide range of tasks, the collaborative nature of MAS allows for far greater reasoning quality and reliability~\cite{du2023improving, li2025knowropesheuristicstrategy}. 

Collaboration strategies in MAS can follow two paradigms.
The first is dynamic coordination, determined at runtime, where agents flexibly communicate intentions, share information, and negotiate task assignments.
The second is a more predictable, workflow-based approach, where intra- and inter-agent interactions are structured as a graph. The graph predefines the execution flow---i.e., code path.
Examples of the latter include prompt-chaining~\cite{wei2022chain}, parallelization and routing~\cite{langgraph-workflows}, self-consistency~\cite{wang2023self}, self-refine~\cite{madaan2023self}, and various combinations thereof. In practice, complex MAS applications adopt a mixture of the two approaches.

\section{Problem definition}\label{sec:motivation}
With MAS continuing to grow in complexity and size, there is increasing consensus towards deploying and integrating MAS with distributed systems~\cite{chaudhry2025compound,kagent,temporal-agents-sdk}.
Powered with standardized communication protocols such as Google's Agent2Agent Protocol (A2A)~\cite{agent2agent-protocol} and Anthropic's Model Context Protocol (MCP)~\cite{model-context-protocol}, \textbf{d}istributed MAS (\textbf{DMAS}) are emerging as a novel trend. We first motivate this trend, then we discuss the limitations of existing programming frameworks in supporting DMAS.

\subsection{Why distributed systems?} 

Akin to microservices, distributing agents into their own services or containers, rather than deploying them as part of a monolith, offers significant architectural and operational benefits. This approach enables heterogeneous runtimes and simplifies dependency management, allowing each team to adopt the agentic framework that best suits their use case without bloating the system with unnecessary libraries. It also promotes agent reuse, where a robust, fault-tolerant deployment of a specific agent can be leveraged across multiple applications.
Security boundaries are strengthened through containerization, which allows fine-grained controls such as seccomp/AppArmor profiles, separate service accounts, and tailored network policies — particularly important when handling Personally Identifiable Information (PII), integrating with external APIs, or managing third-party secrets. 

Furthermore, isolating agents can reduce tail latency by mitigating the impact of failures or model timeouts, ensuring more predictable system responsiveness. If a container crashes or a node fails, only the affected part of the workflow needs to be retried, rather than restarting the entire request. 
In certain scenarios, the retries might be avoided due to the stochastic behavior of LLMs and the inherent adaptability of MAS. For example, when agents coordinate dynamically (\cref{sec:agent}), one may decide to entirely disregard the failed interactions (e.g., due to the crash of a tool or agent container) without compromising the quality of the overall outcome. 

Finally, containers provide resource isolation and enable independent scaling, so similarly with serverless~\cite{bilal23serverless}, agents with diverse CPU, memory, or I/O requirements can be tuned and deployed optimally, avoiding contention and improving overall system performance.



\smallskip
\subsection{Disconnect between programming frameworks and DMAS}
Existing programming frameworks--such as LangGraph~\cite{langgraph}, CrewAI~\cite{crewai},  AutoGen~\cite{autogen}, LlamaIndex~\cite{llamaindex} and Agno~\cite{agno}-- enable programmers to structure applications in workflows and agents, and offer built-in modules for agent coordination, tool integration and memory management.
However, they
tightly couple the application logic with the execution environment by hard-coding communication interfaces. 
Each specific framework implements inter-agent communication via its own modules assuming a monolithic deployment (e.g., message passing in LangGraph or group chats in Autogen).
As a consequence, while it is practical for AI engineers to write MAS applications leveraging today's frameworks, deploying such applications as a distributed, protocol-compliant system is considerably more demanding. It requires substantial manual effort, especially when the deployment environment needs to be changed (e.g., porting a MAS from Kubernetes~\cite{kagent} to Temporal~\cite{temporal-agents-sdk}).
In real cases, practitioners are often puzzled on how to achieve this~\cite{langgraph-containers-so, acp-issue-227,reddit-langchain}. 

\smallskip
\smartparagraph{Example} Consider a LangGraph workflow~\cite{langgraph-workflows}. To deploy it as a distributed system, one would need to:
\begin{enumerate*}[label=\emph{(\roman*)}]
\item decide a partitioning logic to create a distributed graph made of smaller sub-graphs;
\item create separate LangGraph workflow for each sub-graph;
\item stich-the-dots, trying to connect the new sub-graphs while preserving the original orchestration logic;
\item in doing so, implement protocol adapters to translate between LangGraph's communication primitives and A2A/ACP primitives;
\item scaffold deployment-specific infrastructure to run the distributed system on the desired environment--e.g., in Linux containers,  serverless lambdas, Kagent resources~\cite{kagent}, VMs in E2B~\cite{e2b}, or Temporal workers~\cite{temporal-agents-sdk}.
\end{enumerate*}

\smallskip
\smartparagraph{Limitations}
Unfortunately, this manual effort must be repeated for every new communication protocol, changes to the dependencies between agents, and new deployment environment--\emph{i.e.}, it is not a one-time cost.
The problem is exacerbated by the rapid proliferation of different solutions in the field, which forces developers to repeatedly re-engineer their applications to keep up with the latest trends.
Further, it hinders opportunities for optimization. For example, step \emph{(i)} could be optimized based on runtime profiling of communication costs and computational load. 
Similarly, optimization opportunities exist for other design choices, including which communication protocol to use or which runtime environment.
Therefore, automation is fundamental, since one might need to iteratively re-deploy and profile a DMAS for each design choice.



\section{Our approach \& vision}\label{sec:vision}

\begin{figure}[!t]
\includegraphics[width=\linewidth]{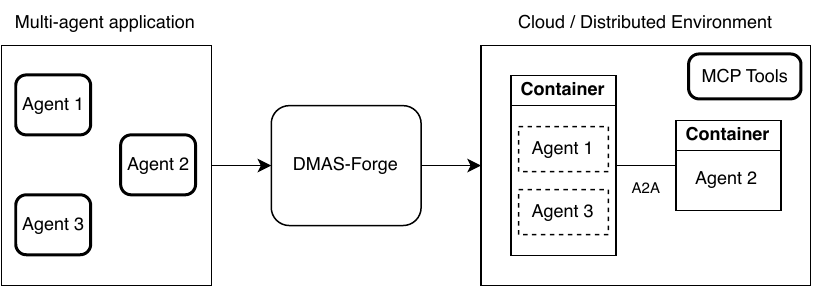}%
\caption{Objective of \sysname.}\label{fig:DMAS-forge}%
\end{figure}

We envision a ``\emph{write-once, deploy-everywhere}'' paradigm, where developers write multi-agent applications once and our framework flexibly ports them to different execution environments, compliant with communication protocols. This vision is summarized in \cref{fig:DMAS-forge}.

\subsection{DMAS-Forge compiler}

We propose a compiler-based approach that enables clean separation of core agent logic from the underlying communication protocols and deployment infrastructure. 
Our key insight is that the core computation model of an AI application is completely orthogonal to how and where the computation is performed. This clean separation allows application developers to provide their structural agentic workflow independently from the deployment choices, and allows developers to plug-and-play any deployment choice in the future. 

Our compiler expects two inputs, highlighted in \cref{fig:code_example}. 
\begin{itemize}[wide, labelwidth=0pt, noitemsep, topsep=0pt, label={}]
\item \emph{The structural agentic workflow (ln.\ref{line:workflow-spec}):} a graph-like computation workflow that includes the various agent implementations with their corresponding tools and the connections between the different agent computations. We follow the computational graph model of an existing AI programming framework, LangGraph, that represents the nodes in the graphs as agents and the edges between the nodes as agent communications.
\item \emph{The deployment specification} (ln.\ref{line:deploy-spec}):
this is the deployment information for the computation graph  including how each agent/tool will run (i.e., process, container, serverless), how different agents/tools will connect and communicate (i.e., choice of protocol), runtime constraints (e.g., hardware type, number of replicas, access policies), and the underlying LLM for each agent.
\end{itemize}

The compiler automatically generates the necessary glue code for deploying the application as described by the provided workflow and the deployment specification. It automatically bakes in the code to ensure that connected agents/tools comply with the user-specified communication protocol. Additionally, depending on the deployment targets, it automatically generates the necessary configuration files, including Dockerfiles and Kubernetes configurations to enforce access policies, ensure the binding to the desired LLM type, etc.

Our compiler-based approach is inspired by previous approaches to flexibly configure microservices~\cite{Blueprint}, and flexibly support distributed deployment of computation graphs in different environments~\cite{tensorflow}.



\section{Prototype}\label{sec:prototype}

\begin{figure}[!t]
\centering
\begin{lstlisting}[frame=single, escapeinside={@}{@}]
import DMASForge/http
import DMASForge/linuxcontainer
import DMASForge/vllm
import DMASForge/openai

@{\textbf{spec = DMASForge.NewSpec}@()

// Deployment specification @\label{line:deploy-spec}@
def DeployAgent(agent):
  http.Deploy(spec, agent)
  linuxcontainer.Deploy(@\textbf{spec}@, agent)

// Structural agentic workflow @\label{line:workflow-spec}@
model = vllm.Model("gpt-4o")
weather = openai.Agent(model, prompt="...")
news = openai.Agent(model, prompt="...")
weather.Connect(news)

// Deployment
DeployAgent(news)
DeployAgent(weather)
\end{lstlisting}
\caption{Two-agent application in \sysname.}
\label{fig:code_example}
\end{figure}

We showcase an initial prototype\footnote{Available at \url{https://github.com/vaastav/DMAS_forge}} in Go language, named \sysname.
To build \sysname, we extend the Blueprint microservices compiler~\cite{Blueprint} to support AI applications. We chose Blueprint as it is compatible with our computational model and it already provides a large array of deployment choices and infrastructure components that can be leveraged.

\begin{table}[t]
\centering
\begin{tabularx}{\linewidth}{lX}
\toprule
\multicolumn{2}{l}{\textbf{Workflow Type Extensions}} \\
\midrule
\codefrag{Agent} & Specifies an Agent \\
\codefrag{ .Connect(agent)} & Connects two agents \\
\codefrag{ .AddTool(tool)} & Adds a tool \\
\codefrag{Model} & Specifies a Model \\
\codefrag{Tool} & Specifies a Tool that an Agent can use \\
\midrule
\multicolumn{2}{l}{\textbf{Wiring API}}\\
\midrule
\codefrag{NewSpec()} & Creates a new DMAS-Forge spec \\
\codefrag{openai.Agent()} & New instance of an openai Agent\\
\codefrag{vllm.Model(name)} & Launches a new model instance \\
\codefrag{NewTool()} & New instance of a tool \\
\codefrag{a2a.Deploy()} & Deploy Agent with A2A\\
\codefrag{mcp.Deploy()} & Deploy Tool using MCP\\
\codefrag{kagent.Deploy()} & Deploy via kagent\\
\bottomrule
\end{tabularx}%
\caption{DMAS-Forge API overview.}%
\label{tab:api}%
\end{table}

\begin{figure*}[t]
    \centering
    \includegraphics[width=0.9\linewidth]{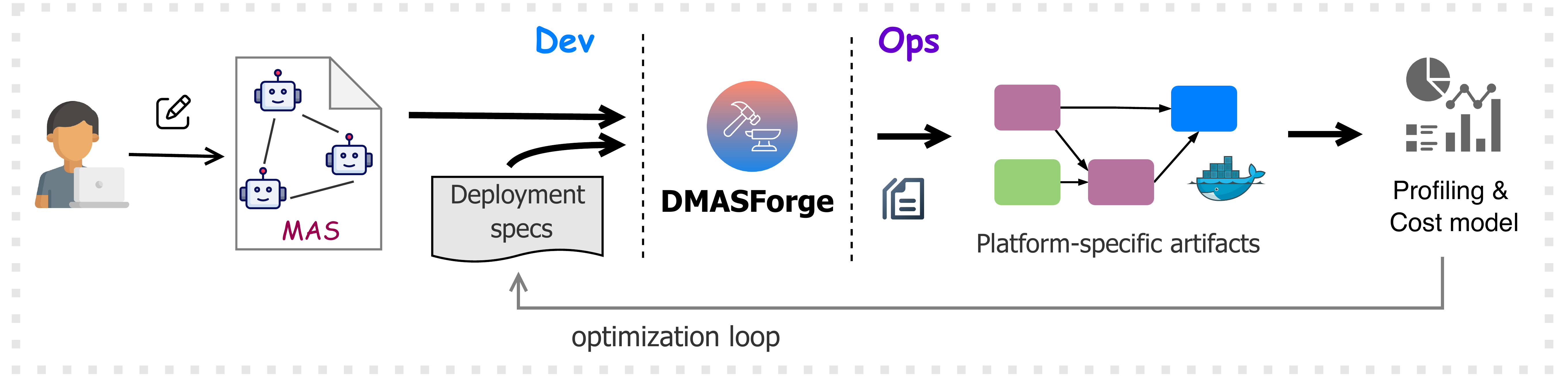}
    \caption{\sysname as enabler of a closed-loop optimization pipeline for MAS deployment.}
    \label{fig:pipeline}
\end{figure*}

\smartparagraph{Programming interfaces}
\autoref{tab:api} shows extensions we add to Blueprint to support the new requirements.
First, we extend Blueprint's \emph{workflow API} to allow users to easily implement their applications as structured workflows, similar to LangGraph.
Second, we offer a new \emph{wiring API}, through which developers can input the deployment specifications to \sysname.

In our prototype, this is achieved by implementing several new Blueprint plugins: (1) An \texttt{Agent} plugin, to transparently create agents and connect them with any OpenAI-compatible model.
(2) A \texttt{vLLM} plugin, to automate model deployment in vLLM~\cite{vllm}.
(3) A \texttt{kagent} plugin, for supporting distributed deployments with \texttt{kagent}~\cite{kagent}, an emerging Kubernetes-native framework for AI agents that provides Custom Resource Definitions (CRDs) for agents, tools and models.

Lastly, we leverage the RPC-over-HTTP Blueprint plugin as an example of inter-agent communication protocol in our prototype.
\cref{fig:code_example} shows a complete example of the use of these plugins to deploy two agents as Linux containers.

\section{Discussion and future work}\label{sec:future_work}

\sysname is a compiler-based framework that transforms multi-agent applications into a distributed deployment with the effort of one click. It targets diverse runtime environments and aims at generating the necessary glue code for  each of them. We presented an initial prototype that supports Linux containers. We plan to extend to other runtime environment, as well as showcasing its benefits for (at least) the following areas.

\smartparagraph{Optimization pipelines}
A key area for improvement would be to streamline the creation of closed-loop optimization pipelines. Applications can be written, deployed, and profiled for communication patterns and resource usage, then automatically restructured for better performance. For example, LangGraph’s communication structures (such as sequential, parallel, or conditional flows) can be optimized at runtime without requiring manual redeployment. 
This pipeline is illustrated in \cref{fig:pipeline}.

This process relies on measuring and modeling communication patterns, performance, and cost-efficiency across different infrastructures and deployment environments. It also involves decisions about whether agents should be co-located or separated into different containers. A key question is whether alternative communication patterns or protocols might better suit the application, given its deployment, latency, and performance constraints.

With these improvements, users no longer need to manually specify the number of replicas or communication protocols. Instead, the framework can make those choices automatically, based on the available infrastructure and performance requirements.

\smartparagraph{Automatic security boundaries}
Another area of future work is to explore means of re-adjusting security policies and determining the least amount of authorization for a DMAS container (or other deployed instance), based on the agents and tools it is running. This is particularly useful when agents are co-located or disaggregated across containers by the optimization pipeline. Automatically determining the security boundaries becomes necessary to minimize the attack surface of each deployed container.




\smartparagraph{Engineering challenges}
Future work should align \sysname with the complete features of current agentic frameworks and extend them. A key open question is whether to build comprehensive agentic capabilities directly into \sysname or to serve as an abstraction layer over existing frameworks. The latter approach would require techniques (e.g., monkey patching) to redirect framework-specific communication primitives to protocol-compliant distributed channels and avoid substantial re-engineering effort.

\balance
\bibliographystyle{ACM-Reference-Format}
\bibliography{references}  

\end{document}